\documentclass[12pt]{iopart}

\usepackage{iopams}
\usepackage[latin9]{inputenc}
\usepackage{graphicx}
\usepackage{esint}

\usepackage{epstopdf}

\begin{document}

\title{The electrostatic interaction of an external charged system with a metal surface: a simplified density functional theory approach}

\author{Iv\'an Scivetti$^1$ and Mats Persson$^{1,2}$}
\address{$^1$ Surface Science Research Centre and Department of Chemistry,
University of Liverpool, Liverpool L69 3BX, UK}
\address{$^2$ Department of Applied Physics, Chalmers University of 
Technology, SE-412 96 G\"oteborg, Sweden}
\ead{scivetti@liverpool.ac.uk} 

\begin{abstract}
As a first step to meet the challenge to calculate the electronic
structure and total energy of charged states of atoms and molecules
adsorbed on ultrathin-insulating films supported by a metallic
substrate using density functional theory (DFT), we have developed a
simplified new DFT scheme that only describes the electrostatic
interaction of an external charged system with a metal surface. This
purely electrostatic interaction is obtained from the assumption that
the electron densities of the two fragments (charged system and metal
surface) are non-overlapping and neglecting non-local exchange
correlation effects such as the van der Waals interactions between the
two fragments. In addition, the response of the metal surface to the
electrostatic potential from the charged system is treated to linear
order, whereas the charged system is treated fully within
DFT. In particular, we consider the classical perfect conductor
(PC) model for the metal response, although our formalism is not
limited to this approximation. The successful computational
implementation of this new methodology and the PC model is exemplified
by the case of a Na$^{+}$ cation outside a metal surface.
\end{abstract}


\maketitle

\section{Introduction}
Recent progress in the studies of single atoms and molecules on
ultrathin-insulating films supported by a metal substrate have opened
up a new frontier in atomic scale science. Not only are these films
able to decouple the electronic states of the adsorbates from the
metal substrate states, but they also provide sufficient tunnelling
current to be able to characterise, manipulate and image the adsorbed
species by a scanning tunnelling microscope. Some important examples
include charge state control of adsorbed species
\cite{RepMeyOlsPer04,OlsPaaPerRepMey07}, imaging of frontier orbitals
\cite{RepMeyStoGouJoa05,RepMeyPaaOlsPer06}, coherent electron-nuclear
coupling in molecular wires \cite{RepLilMey10} and tunnelling-induced
switching of adsorbed molecules \cite{liljeroth,mohn}.  In particular,
polar films such as NaCl bilayers are able to support multiple charge
states of adsorbed species, which can be switched in a controlled
manner by attachment and detachment of tunnelling electrons.
Furthermore, these processes have been suggested to be responsible for
the observed reversible bond formation in molecular switches
\cite{mohn}. To fully exploit the new potential opportunities provided
by these systems, there is a need of theory to unravel the electronic
and geometric structure and the excited state potentials of anionic
and cationic states of the adsorbed species.

Some important advances in this respect have been made by using
Density Functional Theory (DFT) \cite{HK,KS}
calculations~\cite{RepMeyOlsPer04,RepMeyPaaOlsPero5,RepMeyPaaOlsPer06,OlsPaaPerRepMey07,mohn}.
However, the delocalisation error in the current exchange-correlation
functionals and the system size make the calculations very
challenging. In fact, this error in current density
functionals can result in fractional occupation of the adsorbed
species and makes it difficult to identify various multiply charged
states \cite{CohenScienceDFTfailures08}.  Even with these limitations,
important progress has been made by using DFT+U
\cite{anisimov,cococcioni} to correct for the self-interaction behind
the delocalisation error for a Ag atom \cite{OlsPaaPerRepMey07}, despite
it is generally a bit dubious for delocalised states such as for $s$
states of atoms \cite{sit}. There has also been some developments in
constraining the occupancy of adsorbate orbitals, but this requires
that one has to identify the appropriate Kohn-Sham orbital, which can
still be challenging due to the mixing of adsorbate and substrate
states albeit being very small \cite{RepMeyPaaOlsPero5}. However, when
considering, for example, organic molecules adsorbed on an
ultrathin-insulating film supported by a metal surface, the number of
metal atoms and the associated number of metal electrons makes the
calculations to become prohibitively large.

As a first step to try to overcome these limitations, we present a new
simplified DFT method for the calculation of the total energy, ionic
(Hellman-Feynman) forces and electronic structure of a charged system
placed in front of a metal surface. In this method, the charged system
is fully treated using DFT, the interaction between the two systems is
assumed to be electrostatic and the density response of the metal
surface to the electrostatic interaction is treated to linear
order. The proposed method has two main advantages. Firstly, the
computational effort is significantly decreased, since the metal
electron states are not treated explicitly but only implicitly through
their density response to an external electrostatic field, which can
be captured using a simple model. Here, we have explicitly considered
the classical perfect conductor (PC) model for the metal response in
which the screening length by the conduction electrons is assumed to
be zero and the screening charge only resides on the perfect conductor
plane.  Secondly, different charge states of the charged system can be
handled directly using this method.

We have implemented this new methodology in the VASP code~\cite{vasp} using the
simple PC model for the metal response. A critical test of this
implementation is provided by the example of the Na$^{+}$ ion outside
a perfect conductor. Here, the six electrons in the $p$ semi-core of
the Na$^{+}$ ion were included in the calculations. The results show
an excellent agreement between the calculated Hellman-Feynman force on
the ion and the negative gradient of the interaction energy even in
the region close to the perfect conductor plane, where the electron
density of the ion overlaps with the induced charge density at perfect
conductor plane. Furthermore, the small polarisability of the ion
makes the calculated interaction energy to be in close agreement with
the analytical result for a point charge model outside the perfect
conductor.

In applying this method to study charged adsorbates deposited over
ultrathin-insulating films supported by a metal substrate, we need to
extend the PC model by including the interactions between the film and
the metal substrate arising from overlapping electron densities and
van der Waals interactions. We plan to capture these interactions
through a simple parametrised force field between the metal substrate
and the ions in the film, where parameters will be obtained by fitting
the force field to DFT calculations of the forces between the film
(without adsorbates) and the metal substrate. A successful
implementation of this force field should enable us to describe, for
example, the observed reversible bond formation of a molecular switch
induced by a tunnelling electron and hole attachment to form ionic
adsorbate states~\cite{mohn}. The presentation of this scheme and
associated results are deferred to a future publication.

This paper is organised as follows. In the next Section \ref{sec:theory},
we present the theoretical background of the proposed method. First,
the effective energy density functional and
the associated Hellman-Feynman forces for the charged system 
are derived in Subsection \ref{sub:Approx-energy-DF} under
the assumption that the electrons of the charged system and the metal
surface are distinguishable. The total energy of the
metal surface is expanded up to second order in the resulting electrostatic
interaction between the two systems. In the next Subsection \ref{sub:PBC-dipole},
the non-trivial effects on the electrostatics when imposing periodic
boundary conditions are analysed and appropriate dipole corrections
for the total energy and the Hellman-Feynman forces are derived. The
classical perfect conductor model for the metal surface response is
then introduced in the next Subsection \ref{sub:PC-model}. Here, we
also present analytical expressions for the interaction energy and
force in the case of a point charge (Subsection \ref{sub:IntPointCharge}) in the
presence of periodic boundary conditions, and discuss possible refinements
of the perfect conductor model (Subsection \ref{sub:beyondPC}). In
Section \ref{sec:compimpl}, the keys steps in implementing the PC
model in the plane wave code VASP are presented. As a first test of
this methodology, we show the results of the computational implementation
of the PC model for the case of Na$^{+}$ ion outside a PC (Section \ref{sec:Results}) and
finally give some concluding remarks in Section \ref{sec:conc}.

\section{Theoretical framework}
\label{sec:theory}
In this Section we develop a density functional theoretical description
of a charged system placed in front of a metal surface based on the
assumption that the electrons of both fragments 
are distinguishable, so that the interaction between the two systems
is only electrostatic. The corresponding approximation becomes valid
when the electron densities of the two systems are well separated.
We analyse how the system can be treated in a supercell geometry
and derive expressions for the dipole correction of the Kohn-Sham
potential and the Hellmann-Feynman forces. As an explicit model for
the metal surface, we focus on the classical perfect conductor (PC)
model of the metal surface response, but also discuss possible extensions
of this approximation. For the particular case of a point charge, 
we present explicit expressions for the interaction
energy with a perfect conductor. Throughout the paper, we will make use of electrostatic units.

\subsection{Approximate energy density functional\label{sub:Approx-energy-DF}}
We consider a closed, charged system $S$ of electrons and ions outside
a metal surface $M$. In contrast to the system $S$, the metal surface
will be treated as an open system, kept at a constant chemical
potential $\mu$. The electron densities of $S$ and $M$ are assumed to
be non-overlapping and are given by $n_{s}({\bf r})$ and $n_{m}({\bf
  r})$, respectively. In this case the kinetic energy functional of
the non-interacting electrons is additive in the electron densities of
$S$ and $M$. Furthermore, we will neglect any non-local contributions
between $S$ and $M$ such as the van der Waals interaction so that the
exchange-correlation energy functional is also additive in these
electron densities. The total energy density functional for the
combined system is then given by the following expression:
\begin{eqnarray}
E[n_{m},n_{s}]=E_{s}[n_{s}]+E_{m}[n_{m}]- \mu
N_{m}+\int\rho_{m}({\bf r})\phi_{s}({\bf r})d{\bf
  r}\ .\label{eq:EtotApp}
\end{eqnarray}
Here $E_{s}[n_{s}]$ and $E_{m}[n_{m}]$ are the energy density
functionals for the isolated $S$ and $M$, respectively, whereas
$\rho_{m}({\bf r})$ is the charge density of $M$ and $N_{m}$ the total
number of electrons of $M$. The potential $\phi_{s}({\bf r})$ is
defined as the electrostatic potential generated from the total charge
density $\rho_{s}({\bf r})=-en_{s}({\bf r})+\rho_{i}({\bf r})$ of
electrons and ions in $S$ and is given by,
\begin{equation}
\phi_{s}({\bf r})=\int\frac{\rho_{s}({\bf r^{\prime}})d{\bf
    r^{\prime}}}{|{\bf r}-{\bf r^{\prime}}|}\ .\label{eq:phiDef}
\end{equation}
Note that the approximation in Eqn.(\ref{eq:EtotApp}) is still
meaningful for overlapping electron densities. Furthermore, any
neglected non-additive contributions to the kinetic energy and the
exchange-correlation potential including any non-local contribution
between $M$ and $S$, such as the van der Waals interaction, will be accounted by introducing
a force field between $M$ and $S$.

An effective total energy density functional for $S$ outside $M$ in
term of the electron density $n_{s}({\bf r})$ is obtained by
minimising the energy density functional in Eqn.(\ref{eq:EtotApp})
with respect to $n_{m}$ for fixed $n_{s}$ and $\mu$,
\begin{equation}
\frac{\delta E_{m}}{\delta n_{m}({\bf r})}=\mu+e\phi_{s}({\bf
  r})\label{eq:muDef}
\end{equation}
and, since $\mu$ is constant, we conclude that the ground state
electron density of $M$ in the presence of $S$ is a functional of
$n_{s}$, i. e. $n_{m}=n_{m}[n_{s}]$. The change in energy of $M$
induced by the presence of $S$ is defined as,
\begin{eqnarray}
\Delta E_{m}[n_{s}]&=&(E_{m}[n_{m}]-E_{m}[n_{m0}])- \mu(N_{m}-N_{m0}) +\int\rho_{m}({\bf r})\phi_{s}({\bf r})d{\bf
  r}\label{eq:DeltaEDef}
\end{eqnarray}
where $n_{m0}$ and $N_{m0}$ are the unperturbed ground state
electronic density and total number of electrons of $M$,
respectively. Expanding $\Delta E_{m}[n_{s}]$ up to second order in
$\phi_{s}({\bf r})$, one obtains
\begin{eqnarray}
\Delta E_{m}[n_{s}]&=&\int\frac{\delta\Delta
  E_{m}}{\delta\phi_{s}({\bf r})}\phi_{s}({\bf r})d{\bf
  r}+ \frac{1}{2}\int\int\frac{\delta^{2}\Delta E_{m}}{\delta\phi_{s}({\bf
    r})\delta\phi_{s}({\bf r}^{\prime})}\phi_{s}({\bf r})\phi_{s}({\bf
  r}^{\prime})d{\bf r}d{\bf r}^{\prime}.
\label{eq:DeltaEApp} 
\end{eqnarray}

The two terms in Eqn.(\ref{eq:DeltaEApp}) can now be cast in a more
familiar form using the result
\begin{equation}
\rho_{m}({\bf r})=\frac{\delta\Delta E_{m}}{\delta\phi_{s}({\bf
    r})}\ ,\label{eq:RhomRes}
\end{equation}
which follows from the fact that $\Delta E_{m}[n_{s}]$, defined in
Eqn.(\ref{eq:DeltaEDef}), is stationary with respect to variations in
$n_{m}({\bf r})$ for fixed $\phi_{s}$ . Thus, the first term of the
r.h.s in Eqn.(\ref{eq:DeltaEApp}) is determined by the unperturbed
charge density $\rho_{m0}$ of the metal $M$, whereas the second term
is given by the induced charge density $\rho_{ind}({\bf r})$ in $M$,
which, to linear order in $\phi_{s}$, is given by
\begin{equation}
\rho_{ind}({\bf r})=\int\frac{\delta\rho_{m}({\bf
    r})}{\delta\phi_{s}({\bf r^{\prime}})}\phi_{s}({\bf
  r^{\prime}})d{\bf r^{\prime}}\ .\label{eq:rhoindDef}
\end{equation}
Note that the symmetry of the kernel in Eqn.(\ref{eq:DeltaEApp}) and
Eqn.(\ref{eq:RhomRes}) give rise to the following important symmetry
condition for the density response kernel in Eqn.(\ref{eq:rhoindDef}),
\begin{equation}
\frac{\delta\rho_{m}({\bf r})}{\delta\phi_{s}({\bf
    r^{\prime}})}=\frac{\delta\rho_{m}({\bf
    r}^{\prime})}{\delta\phi_{s}({\bf r})}\ .\label{eq:rhoindSymm}
\end{equation}
Finally, we obtain the following effective total energy functional for
$S$ outside $M$ from Eqns.(\ref{eq:DeltaEApp}) and
(\ref{eq:rhoindDef}) as,
\begin{eqnarray}
E_{eff}[n_{s}]\equiv E_{s}[n_{s}]+\Delta E_{m}[n_{s}]= \nonumber \\ E_{s}[n_{s}] +
\int\rho_{m0}({\bf r})\phi_{s}({\bf r})d{\bf
  r}+\frac{1}{2}\int\rho_{ind}({\bf r})\phi_{s}({\bf r})d{\bf
  r}\ .\label{eq:EeffDef}
\end{eqnarray}
The interaction energy between $S$ and $M$ is then simply defined as
\begin{eqnarray}
E_{int} & \equiv & E_{eff}[n_{s}]-E_{s}[n_{s0}]=\label{eq:EintDef}\\ &
& E_{s}[n_{s}]-E_{s}[n_{s0}]+\int\rho_{m0}({\bf r})\phi_{s}({\bf
  r})d{\bf r} + \frac{1}{2}\int\rho_{ind}({\bf
  r})\phi_{s}({\bf r})d{\bf r}\ ,\label{eq:EintExpr}
\end{eqnarray}
where $n_{s0}$ is the unperturbed ground state electron density of
$S$. The contribution from the first two terms in
Eqn.(\ref{eq:EintExpr}) gives the interaction energy from the
polarisation of the system $S$, whereas the contributions from the
third and fourth terms give the interaction energies of $S$ with the
unperturbed surface density and the polarisation of the metal surface,
respectively.

The condition that the chemical potential of $M$ should be constant
imposes an important constraint on the electrostatic potential from
the induced charge densities in $M$ and $S$. From the functional
derivative of Eqn.(\ref{eq:muDef}) it follows 
\begin{equation}
\mu=\frac{\delta T_{0}}{\delta n_{m}({\bf r})}+\frac{\delta E_{xc}}{\delta n_{m}({\bf r})}-e\phi_{m}-e\phi_{s}({\bf r})\ ,\label{eq:muRes}
\end{equation}
where $\phi_{m}({\bf r})$ is the electrostatic potential 
\begin{equation}
\phi_{m}({\bf r})=\phi_{m0}({\bf r})+\phi_{ind}({\bf r})\ \label{eq:phimDef}
\end{equation}
obtained from $\rho_{m}({\bf r})=\rho_{m0}({\bf r})+\rho_{ind}({\bf r})$.
Inserting Eqn.(\ref{eq:phimDef}) in Eqn.(\ref{eq:muRes}), one
finally has 
\begin{equation}
\mu=\frac{\delta T_{0}}{\delta n_{m}({\bf r})}+\frac{\delta E_{xc}}{\delta n_{m}({\bf r})}-e\phi_{m0}-e[\phi_{ind}({\bf r})+\phi_{s}({\bf r})]\ .\label{eq:phimDef2}
\end{equation}

We now assume that the metal surface $M$ is perpendicular to the z-axis and it is
a semi-infinite system that extends to $-\infty$. 
The contribution from the first three terms of the r.h.s. of Eqn.(\ref{eq:phimDef2})
gives the chemical potential in the absence of the perturbation by
$S$ and, since the electron density is unperturbed far inside the
metal surface ($z\rightarrow-\infty$), this contribution is constant.
Therefore, the contribution from the last two terms has to be zero
in this region, 
\begin{equation}
\phi_{s}({\bf r})+\phi_{ind}({\bf r})\rightarrow0\ ,\ z\rightarrow-\infty\ .\label{eq:PhiZero}
\end{equation}

The next step is to determine how the Kohn-Sham (K-S) potential of
$S$ changes in the presence of $M$. Since the K-S potential is determined
by the functional derivative of the electrostatic energy and exchange
correlation energy terms with respect to the electron density, we
need to investigate how these terms are modified. The exchange correlation
energy term is a universal functional of the density and does not
change. On the other hand, according to Eqn.(\ref{eq:EeffDef}),
the electrostatic energy term in $E_{s}[n_{s}]$, is defined here
as, 
\begin{equation}
E_{el}[n_{s}]=\frac{1}{2}\int\rho_{s}({\bf r})\phi_{s}({\bf r})d{\bf r}\ ,\label{eq:EHDef}
\end{equation}
and changes in the presence of $M$ to the effective electrostatic energy 
\begin{eqnarray}
E_{el}^{eff}[n_{s}]&=&E_{el}[n_{s}]+\int\rho_{m0}({\bf r})\phi_{s}({\bf r})d{\bf r}+
\frac{1}{2}\int\rho_{ind}({\bf r})\phi_{s}({\bf r})d{\bf r}\ .\label{eq:EeleffDef}
\end{eqnarray}
Thus, only the electrostatic potential energy term $-e\phi_{s}({\bf r})$
in the K-S potential is modified and is now given by, 
\begin{eqnarray}
-e\phi_{el}^{eff}[n_{s}]&\equiv&\frac{\delta E_{el}^{eff}[n_{s}]}{\delta n_{s}({\bf r})}=-e[\phi_{s}({\bf r})+\phi_{m0}({\bf r})+\phi_{ind}({\bf r})]\ .\label{eq:phieffRes}
\end{eqnarray}
In deriving this result from Eqns.(\ref{eq:DeltaEApp}), (\ref{eq:RhomRes})
and (\ref{eq:rhoindDef}) , we have made use of the symmetry condition
in Eqn.(\ref{eq:rhoindSymm}). Therefore, the only change of the K-S
potential in the presence of $M$ is simply the inclusion of the electrostatic
potential from the metal surface.

The presence of the metal surface will now change the forces ${\bf F}_{I}$
on the nuclei $I$ at positions $R_{I}$ and charges $eZ_{I}$ in
$S$, and this modification will be only through the electrostatic
potential from the metal surface. Since the effective total energy
functional is stationary around the ground state electron density
and only the electrostatic term has an explicit dependence on the
nuclei positions, the forces ${\bf F}_{I}$ are given by 
\begin{equation}
{\bf F}_{I}=-\nabla_{I}E_{el}^{eff}=-\int\frac{\delta E_{el}^{eff}}{\delta\rho_{i}({\bf r})}\nabla_{{\bf {R}_{I}}}\rho_{i}({\bf r})\ d{\bf r}.\label{eq:Fi}
\end{equation}
In a similar manner as for the functional derivative with respect
to the electron density in Eqn.(\ref{eq:phieffRes}), one obtains that
\begin{equation}
\frac{\delta E_{el}^{eff}}{\delta\rho_{i}({\bf r})}=\phi_{s}({\bf r})+\phi_{m}({\bf r}). \label{eq:FuncDerEeff}
\end{equation}
The expression for the forces in Eqn.(\ref{eq:Fi}) can now be cast
in a more familiar Hellman-Feynman form using Eqn.(\ref{eq:FuncDerEeff})
and, since 
\begin{equation}
\nabla_{{\bf R}_{I}}\rho_{i}({\bf r})=-eZ_{I}\nabla\delta({\bf r}-{\bf R}_{I}),
\end{equation}
one obtains, after partial integration, the total force acting on
the ion $I$ 
\begin{equation}
{\bf F}_{I}=-eZ_{I}\nabla[\phi_{s}+\phi_{m}]({\bf R}_{I}).\label{eq:FiRes}
\end{equation}
Therefore, the force on a nuclei is still given by the total
electrostatic field at the nuclei, but now includes the electrostatic
contribution from the metal surface.

\subsection{Periodic boundary conditions and dipole corrections\label{sub:PBC-dipole}}

Since the combined system will be represented in a finite supercell, we
need to discuss the effects of imposing periodic boundary conditions
(PBCs) \cite{ashcroft,kittel}. Firstly, we will introduce PBCs in
the lateral directions along the planar surface of the semi-infinite
metal and state some general key properties of the behaviour of the
electrostatic potentials. In addition, the system will have a net
dipole moment component along the direction perpendicular to the metal
surface and, when imposing PBCs in this direction, one has to introduce
appropriate dipole corrections for the total energy, K-S potential
and the forces on the nuclei as detailed in the following.\\
 In discussing the behaviour of the electrostatic fields and charge
densities in the presence of lateral PBCs, it is convenient to introduce
a plane wave representation over two-dimensional (2D) reciprocal lattice
vectors ${\bf G}$ of the two dimensional surface unit cell $\mathcal{A}$ with area $A$.
The 2D plane wave expansion of a charge density $\rho({\bf r})$ is
here defined as, 
\begin{equation}
\rho({\bf r})=\sum_{{\bf G}}\rho(z,{\bf G})\exp[i{\bf G}.{\bf R}]\label{eq:2DPlaneWave}
\end{equation}
where the plane wave coefficients $\rho(z,{\bf G})$ are given by
\begin{equation}
\rho(z,{\bf G})=\frac{1}{A}\int_{\mathcal{A}}\rho({\bf R},z)\exp[-i{\bf G}.{\bf R}]d{\bf R}\label{eq:2DCoeff}
\end{equation}
and $({\bf R},z)\equiv{\bf r}$. The electrostatic potential from
the charge density can now be obtained using the associated plane
wave representation of the Coulomb kernel, 
\begin{eqnarray}
\frac{1}{|{\bf r}-{\bf r}^{\prime}|}&=&-\frac{2\pi}{A}|z-z^{\prime}|+ \sum_{{\bf G}\neq0}\frac{2\pi}{AG}\exp[-G|z-z^{\prime}|]\exp[i{\bf G}.({\bf R}-{\bf R}^{\prime})]\ 
\label{eq:2DFourierCoulomb}
\end{eqnarray}
where $|{\bf G}|=G$. In particular, the laterally averaged potential, $\bar{\phi}(z)=\phi(z,{\bf G}=0)$
is determined by the laterally averaged density, $\bar{\rho}(z)=\rho(z,{\bf G}=0)$,
as,
\begin{equation}\label{eq:AvPhi}
\bar{\phi}(z)=-2\pi\int|z-z^{\prime}|\bar{\rho}(z^{\prime})dz^{\prime}+\phi_{0}\ .
\end{equation}
Note that $\bar{\phi}(z)$ is only determined up to a constant $\phi_{0}$,
which has to be fixed by the boundary conditions as discussed later.
Finally, the remaining laterally varying part of the potential, which
has contributions only from the non-zero ${\bf G}$ plane wave coefficients,
is given by 
\begin{eqnarray}
\phi^{\prime}({\bf r})\equiv\phi({\bf r})-\bar{\phi}(z)= 
\sum_{{\bf G}\neq0}\frac{2\pi}{G}\int\exp[-G|z-z^{\prime}|]\exp[i{\bf G}.{\bf R}]\rho(z^{\prime},{\bf G})dz^{\prime}, \label{eq:latvarPhi}
\end{eqnarray}
and decays exponentially away from a localised charge distribution.

In the case of a semi-infinite metal surface, the external electric
field from the charged system sets up an electron current towards
the surface and builds up a localised surface charge distribution
and an electric field that opposes the external field. In a stationary
situation, the perpendicular components of the electric field and current
far inside the surface have to be zero. This electric field is determined
by the laterally averaged electrostatic potential $\bar{\phi}(z\rightarrow-\infty)$
and, according to Eqn.(\ref{eq:AvPhi}), it is given by 
\begin{equation}
\bar{E}_{z}(z)=-\frac{\partial\bar{\phi}}{\partial z}(z)\rightarrow2\pi\int\bar{\rho}(z^{\prime})dz^{\prime}\ ,\ z\rightarrow-\infty\label{eq:Ez}
\end{equation}

Therefore, in order for the electric field to be zero far inside in
the metal, the total charge of $S$ and $M$ has to be zero. Since
the surface charge of the unperturbed metal is zero 
\begin{equation}
\int_{\mathcal{A}}\int\rho_{m0}({\bf r})d{\bf R}dz=0\ ,\label{eq:rhm0Charg}
\end{equation}
the total induced surface charge has to be equal in magnitude to the
total charge $Q_{s}$ of $S$ but with opposite sign, 
\begin{equation}
\int_{\mathcal{A}}\int\rho_{ind}({\bf r})d{\bf R}dz=-Q_{s}\ .\label{eq:SurfCharg}
\end{equation}
The energy of the vacuum level for the unperturbed metal is defined
to be equal to zero, $(\bar{\phi}_{m0}(z)\rightarrow0$, $z\rightarrow\infty$), then
\begin{equation}
\bar{\phi}_{m0}(z)=-2\pi\int[|z-z^{\prime}|+z']\bar{\rho}_{m0}(z^{\prime})dz^{\prime}\ .\label{eq:AvPhim0}
\end{equation}
The undetermined constant $\phi_{0}$ in $\bar{\phi}_{s}(z)+\bar{\phi}_{ind}(z)$
can now be determined from the condition in Eqn.(\ref{eq:PhiZero})
that the chemical potential of the metal is constant: 
\begin{equation}
\bar{\phi}_{s}(z)+\bar{\phi}_{ind}(z)\rightarrow0\ ,\ z\rightarrow-\infty\ .\label{eq:phizerores}
\end{equation}
Using Eqns.(\ref{eq:AvPhi}) and (\ref{eq:phizerores}), this condition
gives, 
\begin{eqnarray}
\bar{\phi}_{s}(z)+\bar{\phi}_{ind}(z)= 
-2\pi\int[|z-z^{\prime}|-z^{\prime}][\bar{\rho}_{s}(z^{\prime})+\bar{\rho}_{ind}(z^{\prime})]dz^{\prime}\label{eq:phizeroRes}
\end{eqnarray}

Since the screening length in a metal is typically short on the order
of 1 \AA{}, the surface charge distribution of the semi-infinite metal
surface $M$ is highly localised and the combined system is neutral,
so that the total system can be confined in a supercell $\mathcal{V}$
of volume $V$ by introducing a PBC in the perpendicular direction.
However, the separation of charges between $M$ and $S$ gives rise
to a dipole moment and a long-ranged potential $\bar{\phi}(z)$ that
has to be treated carefully. In addition, one also has to ensure that
the length $L$ of the supercell along the z-direction is large enough
so that the short-ranged part of the potential $\phi^{\prime}({\bf r})$,
Eqn.( \ref{eq:latvarPhi}), is confined within $\mathcal{V}$ .

In order to proceed, we first need to discuss the electrostatics in
a supercell. The electrostatic potential $\phi({\bf r})$ from a
charge distribution $\rho({\bf r})$ is most efficiently obtained
by introducing a three dimensional (3D) plane wave representation
of the densities and electrostatic potential over the 3D reciprocal
lattice vectors ${\bf g}$ of the supercell. The plane wave expansion
and coefficients $\rho({\bf g})$ of a density $\rho({\bf r})$ are
defined here as, 
\begin{equation}
\left\{ \begin{array}{lcl}
\rho({\bf r}) & = & \sum_{{\bf g}}\rho({\bf g})\exp[i{\bf g}.{\bf r}]\\
\rho({\bf g}) & = & \frac{1}{V}\int_{\mathcal{V}}\rho({\bf r})\exp[-i{\bf g}.{\bf r}]d{\bf r\ .}
\end{array}\right.\label{eq:3DPlaneWave}
\end{equation}
The plane wave coefficients of the electrostatic field $\phi({\bf g})$
from the density $\rho({\bf r})$ are obtained from the solution of
the Poisson equation in reciprocal space and are given by, 
\begin{eqnarray}
\phi({\bf g})=\left\{ \begin{array}{ll}
0 & ,\ {\bf g}=0\\
\frac{4\pi}{g^{2}}\rho({\bf g}) & ,\ {\bf g}\neq0
\end{array}\right.\label{eq:3DCoulomb}
\end{eqnarray}
Note that this solution for $\phi({\bf r})$ from a $\rho({\bf r})$
with a net charge corresponds to the electrostatic potential from
$\rho({\bf r})$ compensated with a uniform neutralising background,
$-\rho({\bf g=0})$ inside the supercell. Furthermore, the undetermined
constant of $\phi({\bf r})$ is set so that the average value $\phi({\bf g}=0)$
of $\phi({\bf r})$ is zero.

In the case when the system has a net dipole moment, an artificial
uniform electrical field has to be generated in the supercell in
order to fulfil the PBCs. An efficient solution to this problem was
first proposed by Neugebauer and Scheffler \cite{scheffler} and later
corrected by Bengtsson \cite{bengtsson}. In this approach, a net
dipole moment of a charge distribution $\rho({\bf r})$ (perpendicular to the z-axis)
within the supercell is compensated by introducing a dipole layer
at $z=z_{dip}$, 
\begin{equation}
\rho_{dip}({\bf r})=m\delta^{\prime}(z-z_{dip})
\end{equation}
where the surface dipole moment density is defined as, 
\begin{equation}
m=\frac{1}{A}\int_{\mathcal{V}}\rho({\bf r})zd{\bf r}\ \label{eq:dipA}
\end{equation}
and the resulting dipole potential in the supercell is given by,
\begin{equation}
\phi_{dip}(z)=4\pi m\left[\frac{z}{L}-\frac{1}{2}\right]\ ,\ 0<z<L\ .\label{eq:dippot}
\end{equation}
when $z_{dip}=L$. By correcting the electrostatic potential in the supercell using
this dipole potential, one obtains the dipole-corrected potential
\begin{equation}
\phi_{dipcor}({\bf r})=\phi({\bf r})+\phi_{dip}(z)\ .\label{eq:dipcorrpot}
\end{equation}
This corrected potential in the supercell is now equal, up to a constant,
to the electrostatic potential from $S$ and $M$ in the absence of the corresponding
PBC in the direction perpendicular to the surface.

The dipole-corrected effective electrostatic energy is obtained
by correcting the electrostatic potentials $\phi_{s}({\bf r})+\phi_{ind}({\bf r})$
and $\phi_{m0}({\bf r})$ in $E_{el}^{eff}$ (Eqn. \ref{eq:EeleffDef}) as 
\begin{eqnarray}
E_{el}^{eff,dip} & = & \int_{\mathcal{V}}\rho_{s}({\bf r})[\phi_{m0}({\bf r})+\phi_{dip0}(z)+\phi_{0}]d{\bf r}+ \\
 && \frac{1}{2}\int_{\mathcal{V}}\rho_{s}({\bf r})[\phi_{s}({\bf r})+\phi_{ind}({\bf r})+\phi_{dip1}(z)+\phi_{1}]d{\bf r}, \label{eq:Eldipeff}
\end{eqnarray}
where $\phi_{dip1}(z)$ and $\phi_{dip0}(z)$ are the dipole potentials
from the surface dipole moment densities of $\rho_{s}({\bf r})+\rho_{ind}({\bf r})$
and $\rho_{m0}({\bf r})$, respectively. The constants $\phi_{1}$
and $\phi_{0}$ have to be determined such that the conditions in Eqns.
(\ref{eq:AvPhim0}) and (\ref{eq:phizeroRes}) are fulfilled, corresponding
to 
\begin{eqnarray}
\phi_{1} & = & -\phi_{s}({\bf r}_{in})-\phi_{ind}({\bf r}_{in})-\phi_{dip1}(z_{in})\label{eq:phi1}\\
\phi_{0} & = & -\phi_{m0}({\bf r}_{out})\label{eq:phi0}
\end{eqnarray}
where ${\bf r}_{in}$ and ${\bf r}_{out}$ refer to well inside and
outside of the metal surface, respectively. The dipole corrections
of the K-S potential and the Hellman-Feynman forces in Eqns.(\ref{eq:phieffRes})
and (\ref{eq:FiRes}) are now simply obtained by replacing the electrostatic
potential $\phi_{m}({\bf r})+\phi_{s}({\bf r})$ by the dipole corrected
potential $\phi_{m}({\bf r})+\phi_{s}({\bf r})+\phi_{dip0}(z)+\phi_{dip1}(z)+\phi_{0}+\phi_{1}$. 

Before closing this section, we present the expressions for
the double counting terms used to evaluate the total energy. In general,
the kinetic energy is obtained from the one-electron sum and the double
counting term as, 
\begin{equation}
T_{0}[n_{s}]=\sum_{i:occ}\epsilon_{i}-\int_{\mathcal{V}}n_{s}({\bf r})v({\bf r})d{\bf r}.\label{eq:T0Res}
\end{equation}
Note that adding a constant to the K-S potential, $v({\bf r})$, does
not change the kinetic energy and $\phi_{0}+\phi_{1}$ does not give
a contribution in this case. The electrostatic part of the double
counting term is given by,
\begin{eqnarray}
E_{el}^{DC} =  -\int_{\mathcal{V}}\rho_{es}({\bf r})[\phi_{m}({\bf r})+\phi_{s}({\bf r})+\phi_{dip0}(z)+\phi_{dip1}(z)]d{\bf r} 
\label{eq:EDCDef}
\end{eqnarray}
where $\rho_{es}({\bf r})=-en_{s}({\bf r})$ is the charge density
from the electrons. Adding this term to the dipole-corrected, electrostatic
potential energy in Eqn.(\ref{eq:Eldipeff}), one obtains,
\begin{eqnarray}
E_{el}^{eff,dip}+E_{el}^{DC} = \nonumber \\
 \frac{1}{2}\int_{\mathcal{V}}\rho_{i}({\bf r})\phi_{i}({\bf r})d{\bf r}-\frac{1}{2}\int_{\mathcal{V}}\rho_{es}({\bf r})\phi_{es}({\bf r})d{\bf r}+\label{eq:Ewaldionelectron}\\
 \frac{1}{2}\int_{\mathcal{V}}[\rho_{i}({\bf r})-\rho_{es}({\bf r})-\rho_{ind}({\bf r})]\phi_{dip1}(z)d{\bf r}+\label{eq:ECDCdip1}\\
 \frac{1}{2}\int_{\mathcal{V}}\rho_{ind}({\bf r})[\phi_{i}({\bf r})-\phi_{es}({\bf r})+\phi_{dip1}(z)]d{\bf r}+\label{eq:ECDCind}\\
 \int_{\mathcal{V}}\rho_{i}({\bf r})[\phi_{m0}({\bf r})+\phi_{dip0}(z)]d{\bf r}+\label{eq:ECDCdip0}\\
 (\phi_{0}+\frac{1}{2}\phi_{1}) Q_{s}. \label{eq:ECDCconstant} 
\end{eqnarray}
where $\phi_{i}({\bf r})$ is the ionic potential. Note that in the frozen core approximation $\rho_{es}({\bf r})$
is replaced by the valence charge density and $\rho_{i}({\bf r})$
includes the frozen core charge density.

\subsection{Classical perfect conductor model for the metal surface response\label{sub:PC-model}}

Now we turn to a specific model for the unperturbed charge density
$\rho_{m0}({\bf r})$ of the metal surface and its density response
$\rho_{ind}({\bf r})$ to an external electrostatic field. In the
first application of the proposed scheme, we will use the classical
perfect conductor (PC) model. At the end of this Section, we will discuss
how to go beyond this simplistic model for the metal surface by using
knowledge obtained from previous DFT calculations of the semi-infinite
jellium model of a metal surface, in which the charge of the nuclei
is smeared out to a positive homogeneous background.

The classical PC model is based on a semi-infinite
jellium model, here assumed to occupy the half-space $z<z_{p}$, but
makes the bold assumption that the screening length is zero, that
is, the metallic electrons are able to screen out perfectly any spatial
variation of the electric field inside the metal. This assumption
implies that 
\begin{equation}
\rho_{m0}({\bf r})=0
\end{equation}
and also that $\rho_{ind}({\bf r})$ will be localised at the jellium
edge at $z=z_{p}$, 
\begin{equation}
\rho_{ind}({\bf r})=\sigma_{ind}({\bf R})\delta(z-z_{p}).\label{eq:rhomsurf}
\end{equation}
The surface charge distribution $\sigma_{ind}({\bf R})$ can now be
determined by imposing the condition of Eqn.(\ref{eq:phizerores})
that the electrostatic potential inside the metal should be zero,
\begin{equation}
\phi({\bf r})=0\ ,\ z<z_{p},\label{eq:phimzero}
\end{equation}
and the laterally averaged part of the induced charge density $\bar{\sigma}(z)$
has to be equal to 
\begin{equation}
\bar{\sigma}(z)=\sigma_{ind}({\bf G}=0)=-\frac{Q_{s}}{A}\label{eq:sigavRes}
\end{equation}
where $Q_{s}$ is the total charge of $S$. The laterally varying
part of the surface charge distribution $\sigma^{\prime}({\bf r})$,
as obtained from the 2D plane wave coefficients with non-zero reciprocal
lattice vectors (${\bf G}\neq0$), can be determined from the condition
in Eqn.(\ref{eq:phimzero}). Since $\phi_{s}({\bf r})$ satisfies
the Laplace equation in the metal, the $z$ dependencies of their plane
wave coefficients for ${\bf G}\neq0$ are given by, 
\begin{equation}
\phi_{s}(z,{\bf G})=\phi_{s}(z_{p},{\bf G})\exp[G(z-z_{p})]\ ,\ z<z_{p}\ .\label{eq:phisG}
\end{equation}
Using the plane wave representation of the Coulomb kernel in Eqn.(\ref{eq:2DFourierCoulomb}),
the corresponding coefficients of the induced electrostatic potential
from the PC are given in this region by, 
\begin{equation}
\phi_{ind}(z,{\bf G})=\frac{2\pi}{G}\exp[G(z-z_{p})]\sigma_{ind}({\bf G})\ ,\ z<z_{p}\ .\label{eq:phiindG}
\end{equation}
Using (\ref{eq:phisG}) the condition in Eqn.(\ref{eq:phimzero}) will now be obeyed
if and only if, 
\begin{equation}
\sigma_{ind}({\bf G})=-\frac{G}{2\pi}\phi_{s}(z_{p},{\bf G})\ ,\label{eq:sigG}
\end{equation}
which in a real space corresponds to the following condition on the
laterally varying parts of $\phi_{s}({\bf r})$ and $\sigma({\bf R})$,
\begin{equation}
\sigma_{ind}^{\prime}({\bf R})=-\frac{1}{2\pi}\frac{\partial\phi_{s}^{\prime}}{\partial z}({\bf R},z_{p})\ .\label{eq:latvarsigma}
\end{equation}
Note that the derivations of Eqns.(\ref{eq:sigG}) and (\ref{eq:latvarsigma})
are based on the assumption that the electron densities of $S$ is
not overlapping. In the case of such an overlap, the result in Eqn.(\ref{eq:latvarsigma})
in contrast to the result in Eqn.(\ref{eq:sigG}) does not obey the
symmetry condition for the response kernel in Eqn.(\ref{eq:rhoindSymm}). 
Since in practise a small overlap cannot be avoided we will use
the result in Eqn.(\ref{eq:sigG}) rather than the result in Eqn.(\ref{eq:latvarsigma}).

The induced electrostatic potential $\phi_{ind}({\bf r})$ can now
be obtained outside the PC from the induced surface
charge distribution using Eqns.(\ref{eq:sigavRes}) and (\ref{eq:sigG})
but some care is needed to handle the boundary condition for the laterally
averaged part of the induced potential. The laterally averaged potential
$\bar{\phi}_{s}(z)$ from a charge distribution of $S$ that is well-separated
from the PC is given by, 
\begin{equation}
\bar{\phi}_{s}(z)=\frac{2\pi Q_{s}}{A}(z-\bar{z}_{s})\ ,\ \ z\leq z_{p}\ ,
\end{equation}
where $\bar{z}_{s}$ is the centroid of the charge distribution of
$S$ defined as, 
\begin{equation}
\bar{z}_{s}=\frac{1}{Q_{s}}\int\bar{\rho}_{s}(z)zdz.
\end{equation}
Since according to the condition in Eqn.(\ref{eq:phimzero}), $\phi_{ind}(z_{p})=-\phi_{s}(z_{p})$,
the resulting laterally averaged induced electrostatic potential from
Eqn.(\ref{eq:sigavRes}) in the region outside the PC
is given by, 
\begin{equation}
\bar{\phi}_{ind}(z)=\frac{2\pi Q_{s}}{A}(z+\bar{z}_{s}-2z_{p})\ ,\ z>z_{p}\ .\label{eq:latavphimRes}
\end{equation}

The plane wave coefficients from the laterally varying part of the
induced electrostatic potential in the region outside the perfect
conductor is now obtained directly from Eqns.(\ref{eq:latvarPhi})
and (\ref{eq:sigG}) as 
\begin{equation}
\phi_{ind}(z,{\bf G})=-\phi_{s}(z_{p},{\bf G})\exp[-G(z-z_{p})]\ ,z>z_{p}\ .\label{eq:latvarphimRes}
\end{equation}
Note that according to Eqns.(\ref{eq:latavphimRes}) and (\ref{eq:latvarphimRes}),
the induced electrostatic potential in real space is given outside
the PC by the classical image potential corresponding
to the electrostatic potential from the mirror image of the charge
distribution of $S$ in the plane $z=z_{p}$ as given by, 
\begin{equation}
\phi_{ind}({\bf R},z)=-\phi_{s}({\bf R},2z_{p}-z)\ ,\ \ z>z_{p}\ .
\end{equation}

\subsubsection{Interaction with a point charge\label{sub:IntPointCharge}}

In evaluating the proposed scheme based on a perfect conductor (PC) model, it 
is useful to have the result for the interaction energy $E_{int}(z_{0})$
between the PC and an external point charge $\rho_{s}({\bf r})=Q_{s}\delta({\bf r}-{\bf r}_{0})$,
located at a position ${\bf r}_{0}$ in the supercell.
The interaction energy is obtained from Eqns.(\ref{eq:EintExpr})
and (\ref{eq:Eldipeff}) and can be decomposed into one contribution
arising from the lateral averaged potential, $\bar{E}_{int}(z_{0})$,
and one contribution arising from the laterally varying part of the
potential, $E_{int}^{\prime}(z_{0})$, 
\begin{eqnarray}
\bar{E}_{int} & = & \frac{Q_{s}}{2}[\bar{\phi}_{ind}(z_{0})+\phi_{dip1}(z_{0})+\phi_{1}]\label{eq:EintAvDef}\\
E_{int}^{\prime} & = & \frac{Q_{s}}{2}\phi_{ind}^{\prime}(z_{0})\label{eq:EintVarDef}
\end{eqnarray}
where, according to Eqns.(\ref{eq:dippot}) and (\ref{eq:phi1}),
the dipole corrections terms are determined by, 
\begin{eqnarray}
& &\phi_{dip1}(z) =\frac{4\pi Q_{s}}{A}(z_{0}-z_{p})[\frac{z}{L}-\frac{1}{2}]\label{eq:PhiDip1Res}\\
& & \phi_{1}  =  -[\bar{\phi}_{ind}(z_{p})+\bar{\phi}_{s}(z_{p})+\phi_{dip1}(z_{p})]. \label{eq:PhiDip1ConRes}
\end{eqnarray}
Note that in the absence of perpendicular periodic boundary conditions
$\bar{\phi}_{ind}(z_{0})$ is given by Eqn.(\ref{eq:latavphimRes})
and $\phi_{dip1}(z_{0})+\phi_{1}=0$. The laterally averaged electrostatic
potentials are given in the supercell by, 
\begin{eqnarray}
\bar{\phi_{s}}(z) = \frac{2\pi Q_{s}}{A}[-|z-z_{0}|+\frac{(z-z_{0})^{2}}{L}+\frac{L}{6}] & & \label{eq:latavphisPBC}\\
\bar{\phi}_{ind}(z) =  -\frac{2\pi Q_{s}}{A}[-|z-z_{p}|+\frac{(z-z_{p})^{2}}{L}+\frac{L}{6}]. & &  \label{eq:latphiindPBC}
\end{eqnarray}
Inserting these results into Eqns.(\ref{eq:PhiDip1ConRes}), one obtains
that 
\begin{eqnarray}
\bar{\phi}_{ind}(z_{0})+\phi_{dip1}(z_{0})+\phi_{1}  = 
 \frac{4\pi Q_{s}}{A}[(z_{0}-z_{p})-\frac{L}{12}]\ ,\label{eq:PhiIndSuper}
\end{eqnarray}
and the laterally averaged interaction energy is finally given by,

\begin{equation}
\bar{E}_{int}(z_{0})=\frac{2\pi Q_{s}^{2}}{A}[(z_{0}-z_{p})-\frac{L}{12}]\ .\label{eq:EintAvRes}
\end{equation}
Note that this result differ from the result obtained in the absence
of perpendicular periodic boundary conditions, 
\begin{equation}
\bar{E}_{int}(z_{0})=\frac{2\pi Q_{s}^{2}}{A}(z_{0}-z_{p})\ .\label{eq:EintAvInftyRes}
\end{equation}
by the extra term
\begin{equation}
\frac{Q_{s}}{2}\bar{\phi}_{s}(z_{0})=-\frac{\pi Q_{s}^{2}L}{6A}\ .\label{eq:PBCselfenergy}
\end{equation}
Furthermore, it is noteworthy that this term diverges when $L\rightarrow\infty$
for fixed $A$. The interaction energy $\bar{E}_{int}$ in Eqn.(\ref{eq:EintAvInftyRes})
is nothing else than the electrostatic energy of a parallel plate
capacitor where the two plates, each with an area $A$, are separated
by the distance $z_{0}-z_{p}$ and having charges $-Q_{s}$ and $Q_{s}$.
This repulsive interaction energy vanishes in the limit $A\rightarrow\infty$.

In the case of the laterally varying part, $\phi_{ind}^{\prime}({\bf r})$
is obtained from $\rho_{s}(z,{\bf G})=\frac{Q_{s}}{A}\delta(z-z_{0})$ and Eqns.(\ref{eq:latvarPhi})
and (\ref{eq:latvarphimRes}) as,
\begin{eqnarray}
\phi_{ind}^{\prime}({\bf r})= -\frac{2\pi Q_{s}}{A}\sum_{{\bf G}\neq0}\frac{\exp[-G(z+z_{0}-2z_{p})]}{G}\exp\left[i{\bf G.({\bf R}-{\bf R}_{0})}\right] \nonumber
\end{eqnarray}
when the induced potential is well-localised within the supercell
corresponding to $2\pi(L-z_{0})/L_{\|}\gg1$ and $2\pi z_{0}/L_{\|}\gg1$, with $L_{\|}=L_{x,y}$ the 
length of the supercell in the $x$ and $y$ directions, parallel to the PC plane.
The resulting laterally varying part of the interaction energy, $E_{int}^{\prime}$, 
in Eqn.(\ref{eq:EintVarDef}) is then given directly by,
\begin{equation}
E_{int}^{\prime}(z_{0})=-\frac{2\pi Q_{s}^{2}}{A}\sum_{{\bf G}\neq0}\frac{\exp[-2G(z_{0}-z_{p})]}{2G}.\label{eq:EintvarRes}
\end{equation}
This part of the interaction energy is always attractive and crosses
over into the classical image potential, 
\begin{equation}
E_{int}(z_{0})=-\frac{Q_{s}^{2}}{4(z-z_{0})}\\ ,\label{eq:ClassImage}
\end{equation}
of the point charge in the limit $A\rightarrow\infty$.

The Hellman-Feynman force $F_{int}(z_{0})$ on the point charge is
determined by the dipole corrected electric field at the point charge as, 
\begin{eqnarray}
{\bf F}_{int}(z_{0})=-Q_{s}\nabla[\bar{\phi}_{ind}(z_{0})+
\phi_{ind}^{\prime}({\bf r}_{0})+\phi_{dip1}(z_{0})+\phi_{1}]\label{eq:FpointChargeDef}
\end{eqnarray}

Using the expressions for $\bar{\phi}_{ind}(z_{0})$ and the plane
wave coefficients $\phi_{ind}^{\prime}({\bf r})$ in Eqns.(\ref{eq:latavphimRes}),
(\ref{eq:latvarphimRes}) and (\ref{eq:PhiDip1Res}), one obtains directly, 
\begin{eqnarray}
{\bf F}^{z}_{int}(z_{0})=-\frac{2\pi Q_{s}^{2}}{A}\left(1 + \sum_{{\bf G}\neq0}\exp[-2G(z_{0}-z_{p})]\right) \label{eq:FpointChargeRes}
\end{eqnarray}

Note that this result for the force is consistent with the result
${\bf F}^{z}_{int}(z_{0})=-\nabla E_{int}(z_{0})\hat{z}$ obtained from
the interaction energy $E_{int}(z_{0})=\bar{E}_{int}(z_{0})+E_{int}^{\prime}(z_{0})$
in Eqns.(\ref{eq:EintAvRes}) and (\ref{eq:EintvarRes}).

\subsubsection{Beyond the perfect conductor model\label{sub:beyondPC}}

The first step in going beyond the prefect conductor approximation for the metal
surface is to account for the non-zero screening length of the conduction
electrons. Important information and concepts about the behaviour
of this response have been drawn in the pioneering DFT studies of
the semi-infinite jellium model based on the LDA by Lang and Kohn\cite{kohn_lang}.
They showed that there is a spill-out of electrons into the vacuum
region that creates an extended dipole layer with a surface charge
density $\rho_{m0}(z)$. The dipole moment of this distribution was
shown to determine the surface contribution to the work function.
This charge contribution is not accounted for in the PC
model. From their calculations of the response of the semi-infinite
jellium to an external homogeneous electric field, they showed that
the classical image plane is located at the centroid $z_{im}$ of
the induced density and not at the jellium edge as in the perfect
conductor model. However, this effect is easily accounted for in the
PC model by choosing $z_{p}=z_{im}$.

The static response of the semi-infinite jellium to a laterally varying
external potential can be characterised by a wave-vector dependent
reflection coefficient $g(G)$\cite{LiebRefl}. The 2D plane wave
coefficient of the induced electrostatic potential outside the induced
metal density is given by, 
\begin{eqnarray}
\phi_{ind}(z,{\bf G}) & = & -g(G)\phi_{ext}({\bf G})\exp[-Gz] \label{eq:reflDef}
\end{eqnarray}

where $\phi_{ext}({\bf G})$ is the 2D plane wave coefficient of the external electrostatic
potential. For example, the interaction energy of a single point charge $Q_{s}$
at $z=z_{0}$ with the metal surface becomes, 
\begin{equation}
E_{int}(z_{0})=-\frac{Q_{s}^{2}}{2}\intop_{0}^{\infty}dG\exp(-2Gz_{0})g(G)\ .\label{eq:EintRefl}
\end{equation}
Since the PC model assumes perfect screening for all parallel wave
vectors, $g(G)=\exp(2gz_{p})$, according to Eqn.(\ref{eq:latvarphimRes})
where $z_{p}$ gives the position of the PC plane and
Eqn.(\ref{eq:EintRefl}) reduces to the classical image potential in
Eqn.(\ref{eq:ClassImage}). Calculations by the stabilised jellium
model have shown that PC result for $g(G)$ is an excellent approximation
for $G$ up to $G_{c}=$0.8 \AA{}$^{-1}$ in the range of 2-4 a$_{0}$ for
the electron gas density parameter $r_{s}$\cite{LiebSat}. According
to Eqn.(\ref{eq:EintRefl}), this suggests the PC model is still a
good approximation for the image potential down to distances of about
$1/2G_{C}=0.6$ \AA{}.

\section{Computational details and implementation \label{sec:compimpl}}

The proposed scheme that is based on a perfect conductor model has
been implemented in the plane wave code VASP\cite{vasp}. The key
quantity to compute is the induced electrostatic potential $\phi_{ind}({\bf r})$,
which determines how the total energy, K-S potential and Hellmann-Feynman
forces change in the presence of the PC. The first
step is to generate $\bar{\sigma}_{ind}$ from the total charge of
$S$, Eqn.(\ref{eq:sigavRes}), and the Fourier components of $\sigma_{ind}^{\prime}({\bf R})$
from Eqn.(\ref{eq:sigG}). The induced density of charge $\rho_{ind}({\bf r})=\sigma_{ind}({\bf R})\delta(z-z_{p})$
was then represented at a plane of grid points corresponding to the
position of the PC plane, from which $\rho_{ind}({\bf g})$
was generated by the standard routine in VASP based on Eqn.(\ref{eq:3DCoulomb}).
The surface dipole moment $m$ that determines
$\phi_{dip}$ through Eqn.(\ref{eq:dippot}) was obtained from Eqn.(\ref{eq:dipA}).
The dipole correction term in Eqn.(\ref{eq:ECDCdip1}) is computed
from the standard dipole correction subroutine by $\rho_{ind}$
to $\rho_{es}$. Finally, the computation of energy term of Eqn.(\ref{eq:ECDCind}) is
carried out in reciprocal space.

As a simple illustration and test of the proposed scheme, we have
considered the Na$^{+}$ ion outside the PC. The electron-ion
interaction was described by the projector augmented wave method \cite{paw}.
The six electron in the $p$ semi-core states were treated as valence
electrons. The electronic exchange and correlation effects were treated
within the PBE version \cite{pbe} of the generalised gradient approximation.
The plane wave cut-off energy was set to the standard value of 400
eV and the Brillouin zone was sampled by a 3x3x1 $k$-points.

\section{Results\label{sec:Results}}

\begin{figure}
\centering \includegraphics[scale=0.75]{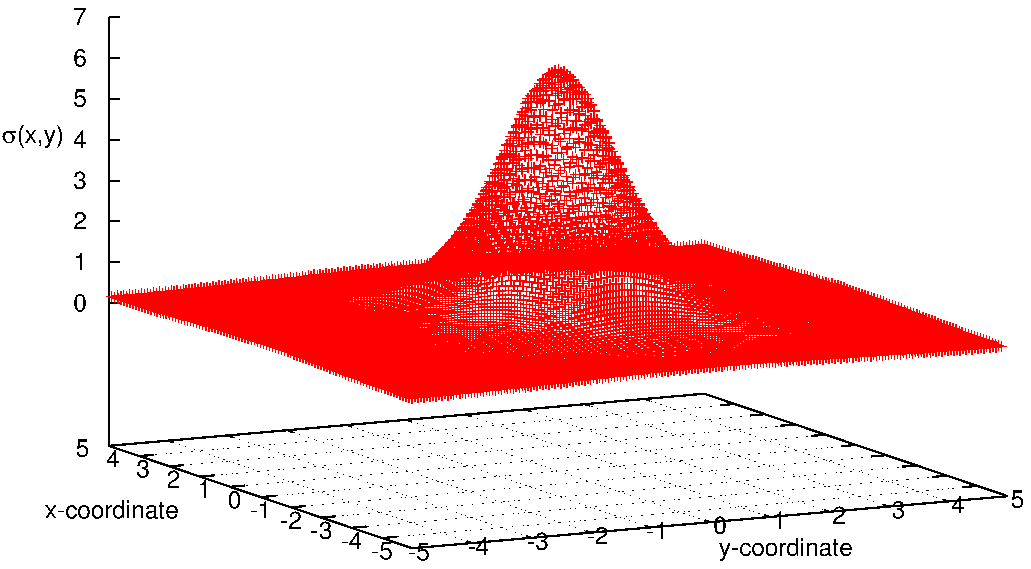}
\includegraphics[scale=0.75]{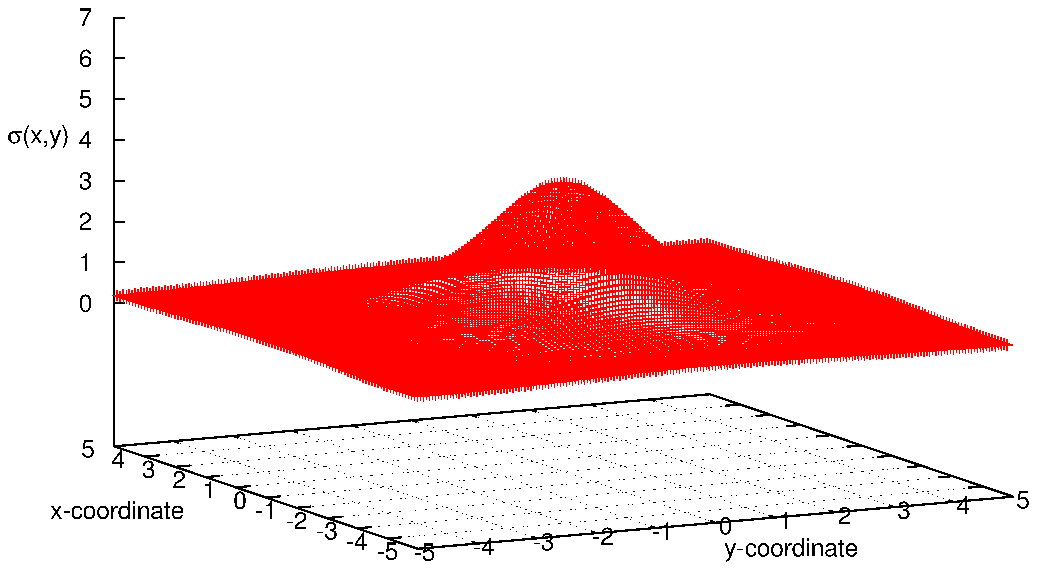}
\caption{Induced surface charge density at the perfect conductor plane for
Na$^{+}$ ion placed at 1.5 \AA{} (Upper panel) and 2.0 \AA{} (Lower panel) away
from the perfect conductor. The size of the surface unit cell
is 10 \AA{} x 10 \AA{} and the height of the supercell is 20 \AA{}.}
\label{fig:ind_charge} 
\end{figure}

The behaviour of the induced surface electron density, $\sigma_{ind}({\bf R})$,
at two different distances of the Na$^{+}$ ion from the perfect conductor
plane is shown in Fig. \ref{fig:ind_charge}, for a supercell
with transversal area $A=100$ \AA{}$^{2}$ and a height L of 20 \AA{}.
It is evident that the induced electron density becomes laterally
more extended as the ion separates from the surface but the net total
charge remains constant and is equal to $-e$. In other words, the
contributions from plane-wave coefficients $\sigma({\bf G}\neq0)$
decrease with the distance and $\sigma_{ind}({\bf R})$ becomes practically
uniform when this distance is sufficiently large.

\begin{figure}
\centering \includegraphics[scale=0.115]{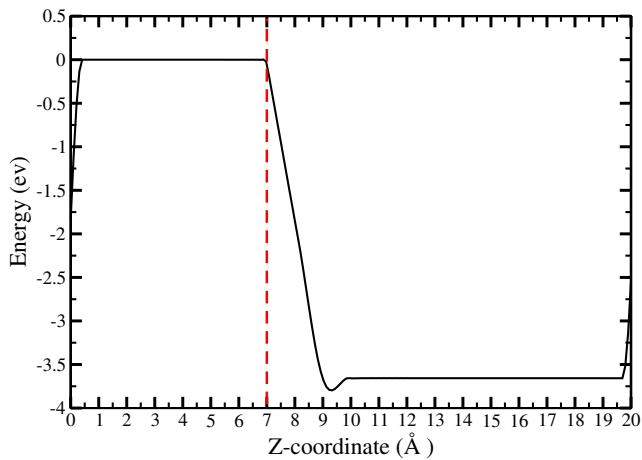}
\caption{Laterally averaged electrostatic potential, $\bar{\phi}(z)=\bar{\phi}_{s}(z)+\bar{\phi}_{m}(z)+\phi_{dip}(z)+\phi_{1}$,
as a function of the $z$ coordinate for a Na$^{+}$ ion placed 2
\AA{} away from the perfect conductor plane located at $z=7$ \AA{}.
The dipole layer is located at $z$ = 20 \AA{}. Same supercell as
in Fig. \ref{fig:ind_charge}.}
\label{fig:pot} 
\end{figure}

The characteristic behaviour of the laterally averaged,
dipole-corrected electrostatic potential,
$\bar{\phi}(z)=\bar{\phi}_{s}(z)+\bar{\phi}_{ind}(z)+\phi_{dip}(z)+\phi_{1}$,
is illustrated in Fig. \ref{fig:pot} as a function of the $z$
coordinate when the Na$^{+}$ ion is placed 2 \AA{} outside the perfect
conductor plane located at $z=7$ \AA{}. Again, the supercell is the
same as in Fig. \ref{fig:ind_charge}. The potential is constant and
equal to zero inside the perfect conductor as dictated by the
condition in Eqn.(\ref{eq:phimzero}). The discontinuous change in
slope at the perfect conductor plane (located at $z=7$\AA) is induced
by the laterally averaged surface charge density $\bar{\sigma}$. In
the region between the perfect conductor plane and the ion (placed at
$z=9$\AA), the slope is constant corresponding to a constant electric
field. The spatial extension of the charge distribution of the ion is
reflected by the deviation from this linear behaviour.  The calculated
value of the slope in the potential in this region is 1.81 eV/\AA{},
which is precisely the value of $4\pi\bar{\sigma}/\epsilon_{0}$ = 1.81
eV/\AA{} for the z-component of the electrical field inside a
capacitor with plates having opposite surface charge densities of
$\bar{\sigma}=e/A=0.01$ e\AA{}$^{-2}$, since $A=100$ \AA{}$^{2}$.  At
the dipole layer located at $z=$ 20 \AA{} the potential makes a jump,
so that the potential is periodic across the supercell boundaries at
$z=0$ and 20 \AA{}. Note that the representation of the surface charge
on a single plane of grid points give rise to a well-behaved
electrostatic potential in contrast to the dipole layer, which has to
be smeared out in the $z$ direction in order to damp out Gibbs
oscillations.

\begin{figure}
\centering \includegraphics[scale=0.11]{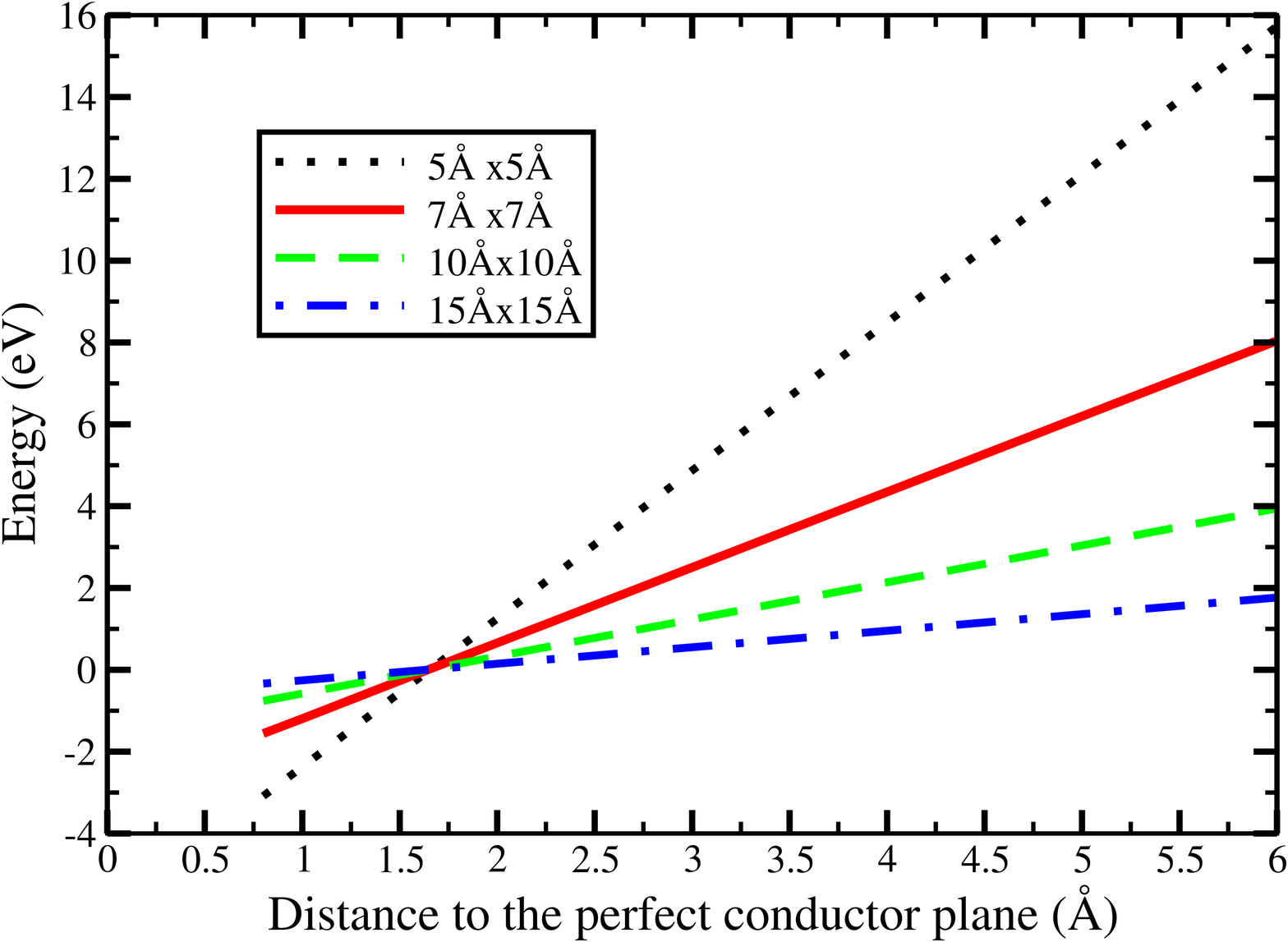}
\includegraphics[scale=0.11]{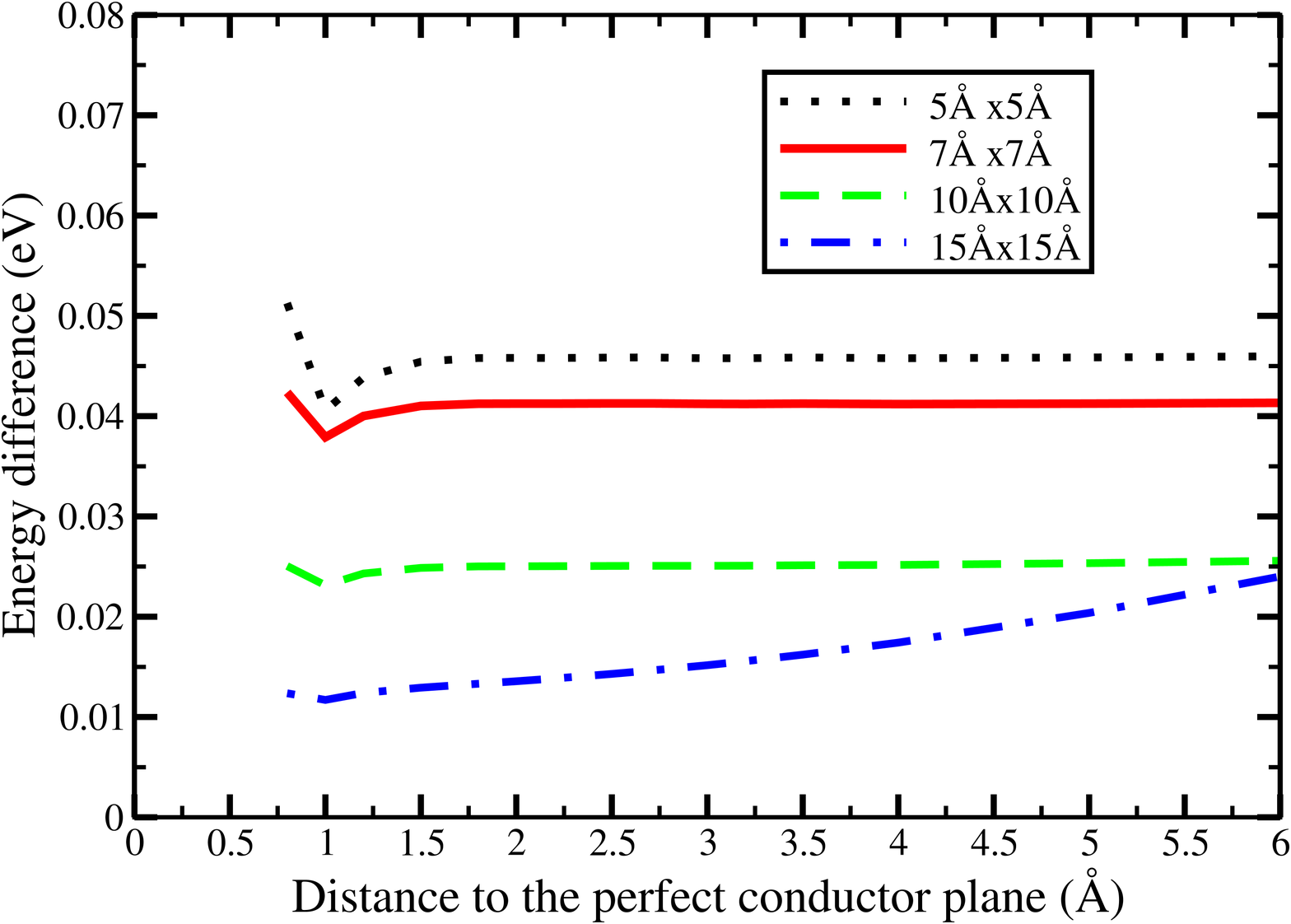}
\caption{(Upper panel) Calculated contribution, $\bar{E}_{int}$, to the interaction
energy from the laterally averaged part of the induced surface charge
density, $\bar{\sigma}$, as a function of the distance to the perfect
conductor plane for the different sizes of the surface unit
cell. (Lower panel) Calculated energy difference with respect to
the point charge case result of Eqn.(\ref{eq:EintAvRes}). The height of
the supercell is 20 \AA{}.}
\label{fig:EG0} 
\end{figure}

\begin{figure}
\centering \includegraphics[scale=0.11]{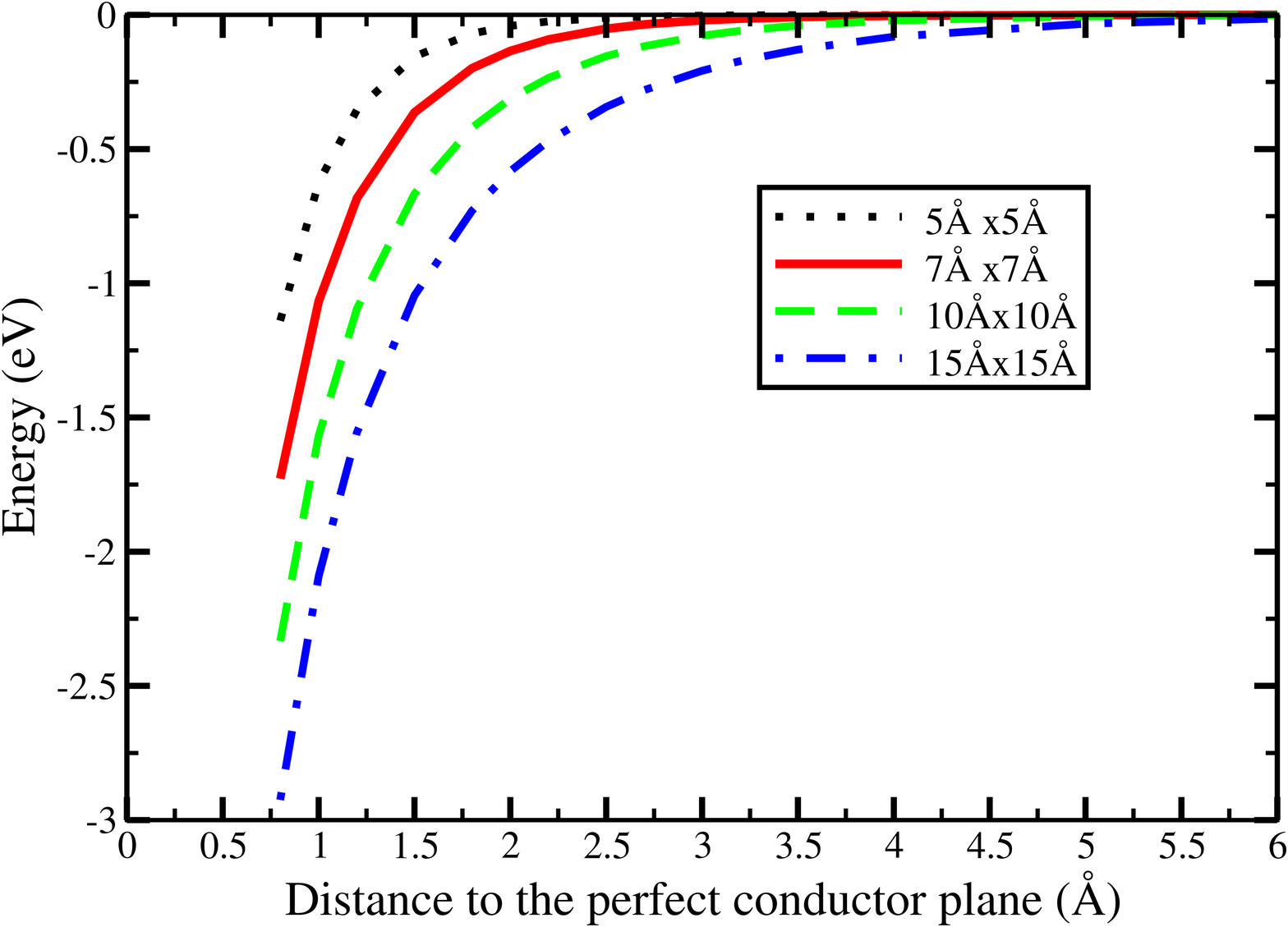}
\includegraphics[scale=0.11]{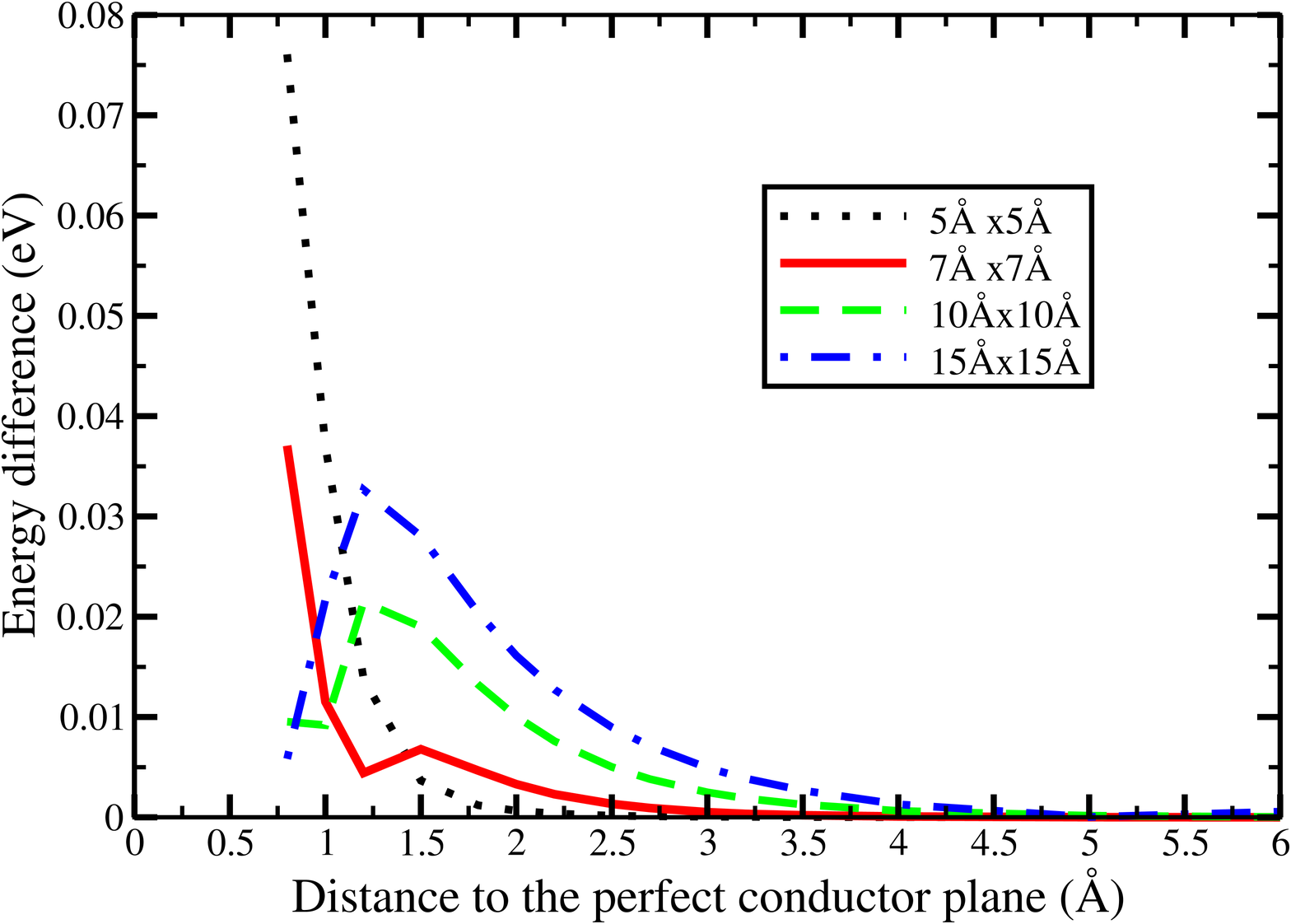}
\caption{(Upper panel) Calculated contribution, $E_{int}^{\prime}$, to
  the interaction energy from the laterally varying part of the
  induced surface charge density, $\sigma^{\prime}({\bf R})$, as a
  function of the distance of the ion to the perfect conductor plane
  for the different sizes of the surface unit cell.(Lower panel)
  Calculated energy difference with respect to the point charge model
  of Eqn.(\ref{eq:EintvarRes}).  The height of the supercell is 20
  \AA{}.}
\label{fig:EneG0} 
\end{figure}

To facilitate the discussion of the behaviour of the calculated interaction
energy, $E_{int}$, between the Na$^{+}$ ion and the PC
as defined in Eqn.(\ref{eq:EintExpr}), we have decomposed this
energy into the contributions $\bar{E}_{int}$ and $E_{int}^{\prime}$
from the laterally averaged, $\bar{\sigma}_{ind}$, and the laterally
varying surface charge density, $\sigma_{ind}^{\prime}({\bf R})$,
respectively. Note that $E_{s}[n_{s0}${]} in Eqn.(\ref{eq:EintExpr})
has been obtained for the ion in the supercell with a uniform neutralising
background. The two contributions
$\bar{E}_{int}$ and $E_{int}^{\prime}$ are shown in Figs. \ref{fig:EG0}
and \ref{fig:EneG0} for different surface areas of the supercell.
Since we shall compare DFT results for the Na$^{+}$ with the analytical
results for a point charge in Eqns.(\ref{eq:EintAvRes}) and (\ref{eq:EintvarRes}),
we need to avoid a significant overlap between the electron density
of the ion and the induced electron density at the perfect conductor
plane. With this, we shall only consider distances of the Na$^{+}$ from the
perfect conductor plane larger than 0.8 \AA{}.

As shown in Fig. \ref{fig:EG0} (Upper panel), the contribution from the uniform
surface charge distribution results in a linear variation of the interaction
energy with the ion distance to the PC surface. This
linear behaviour and the decrease in magnitude of the associated force
(given by the negative gradient) with increasing transversal area
is in agreement with the corresponding result of Eqn.(\ref{eq:EintAvRes})
in the point charge model. The relative difference between the latter
result and the computed interaction energy is smaller than the 2\%,
as shown in Fig. \ref{fig:EG0} (Lower panel). These small differences can
in part be attributed to the polarisation of the semi-core of the
Na$^{+}$and at shorter distances in part the spatial extension of
the charge distribution of the ion. Note that $\bar{E}_{int}$ for
the different $A$ do not cross at the perfect conductor plane but
at a distance of about 1.75 \AA{} outside the perfect conductor plane
due to the extra energy term of Eqn.(\ref{eq:PBCselfenergy}). In fact
this term gives that the crossing is located at $L/12$, which is
in excellent agreement with the result in Fig. \ref{fig:EG0} (Upper panel).

In contrast to the laterally average interaction energy $\bar{E}_{int}$,
the contribution to the interaction energy from the laterally varying
part $E_{int}^{\prime}$ is always attractive, becomes more dominant
with increasing surface area and decay rapidly with the distances
from the perfect conductor, as shown in Fig. \ref{fig:EneG0} (Upper panel).
This result is in close agreement with the result in Eqn.(\ref{eq:EintvarRes}),
obtained from the corresponding contribution in the point charge model,
as shown by the difference between these two results in Fig. \ref{fig:EneG0} (Lower panel).
In fact, relative energy differences between DFT and the point charge
case are smaller than 2\%. In the limit of infinite surface area, the
interaction tends to the classical image interaction.

\begin{figure}
\centering \includegraphics[scale=0.115]{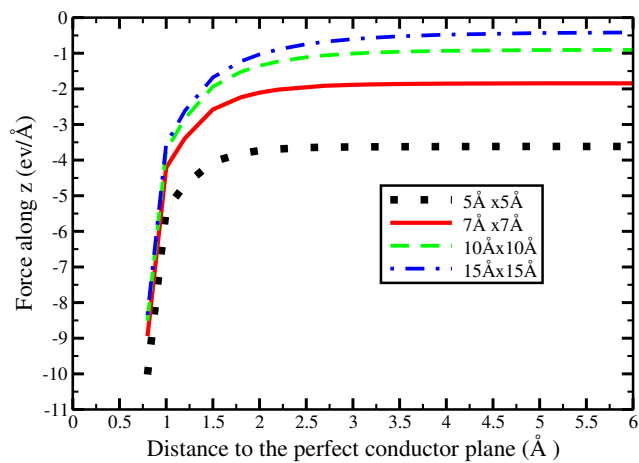}
\caption{Calculated Hellman-Feynman force on the Na$^{+}$ ion as a function
the distance, $z_{0}$, from the perfect conductor plane for the different
sizes of the surface unit cell. The height of the super cell is 20 \AA{}. 
Differences with respect to the point charge case are lower than 2.5\%.}
\label{force_DFT} 
\end{figure}

We now present the calculated Hellmann-Feynman force along the $z$
direction in Fig. \ref{force_DFT}. The force is always attractive
and becomes constant for large distances of the ion from the perfect
conductor plane. As the ion approaches to perfect conductor,
the effect of the laterally varying components of the induced charge
density become increasingly more important, thus leading to an increased
attraction. Clearly, the magnitude of this attraction increases with
the transversal area. In the limit of infinite surface area, the induced
force will tend to the classical force exerted by the image charge.
We observe that the differences with respect to the point charge case are smaller than 2.5\%. 

\begin{figure}
\centering \includegraphics[scale=0.115]{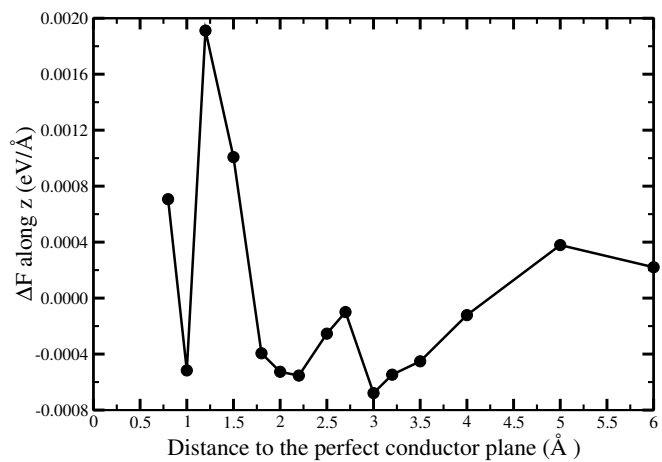}
\caption{Test of the consistency between the Hellman-Feynman force on
  the ion and the force from the gradient of the total energy. $\Delta
  F$ is the difference between these two forces and values are smaller
  than the 0.06\% of the computed forces. Same supercell as in
  Fig. \ref{fig:ind_charge}.}
\label{fig:SCforces} 
\end{figure}

Finally, as a critical test of the implementation of the perfect
conductor model, we analyse the consistency between the computed
Hellman-Feynman force on the ion and the gradient of the total
energy. In Fig. \ref{fig:SCforces}, we show the difference between the
force, $F_{z}$, computed along the $z$ direction using
Eqn.(\ref{eq:FiRes}) and the numerical derivative of the interaction
energy, $-\partial E_{int}/\partial z_{0}$, as a function of the
Na$^{+}$ distance to the perfect conductor. Ideally, this difference
should be zero but, due to numerical errors, small deviations are
always expected.  In fact, errors are essentially smaller than the
0.06\% of the computed forces, showing that sufficient numerical
consistency has been achieved.  This level of self-consistency is also
obtained for smaller distances where we have a more substantial
overlap of the electron densities.  Note that this high degree of
consistency even in the case of overlapping electron densities is
guaranteed by using the density response in Eqn.(\ref{eq:sigG}) that
obeys the symmetry condition in Eqn.(\ref{eq:rhoindSymm}).

The above results suggest that we have derived and implemented a DFT
method for a charged system in front of a perfect conductor. Small
differences with respect to the the point charge case could be attributed
to the differences in the spatial extension of the densities of charge
and that the point charge model does not include any polarisation,
which will be inevitably present when placing atoms or molecules under
the electrical field generated by the induced potential.

We believe that the use of this new DFT methodology becomes particularly
convenient since it allows to the possibility of computing DFT problems
of different charge states in a controllable way, that is, by defining
(at will) the amount of charge transfer between the system and the
perfect conducting plane. However, we have shown that the use of the
perfect conductor approximation only induces attractive interactions,
which would make the system to move closer and closer to the perfect
conductor plane. This represent a limitation if one aims to make use
of this DFT methodology to simulate realistic problems involving metallic
surfaces. In fact, the metallic surface will exert repulsive forces
for sufficiently close distances, thus avoiding the system to collapse
with the surface \cite{pauli}. This lack of repulsion is a direct
consequence of having considered only the electrostatic interactions
in the approximate energy functional of Eqn.(\ref{eq:EtotApp}).
In a following publication \cite{IS_MP}, we propose a new procedure
to surmount this limitation.

\section{Concluding Remarks\label{sec:conc}}
To the final purpose of developing a simplified density functional
theory (DFT) method for treating charged atoms and molecules on an
ultrathin-insulating film supported by a metal substrate, we have
presented a new approximate DFT methodology for the calculation of the
total energy, ionic (Hellman-Feynman) forces and electronic structure
of a charged system placed in front of a metal surface. In this new
methodology, the electron densities of the metal surface and the
charged system are assumed to be non-overlapping and there is only an
electrostatic interaction between these two fragments.  Whereas the
charged system is treated fully using DFT, the metal surface is
approximated within linear response, corresponding to an expansion of
the energy of the metal surface to second order in the electrostatic
potential from the charged system. In particular, we have carried out
a careful analysis of the effect of periodic boundary conditions and
derived appropriate dipole corrections for the total energy and the
ionic forces. The proposed method have two main advantages.  First,
the metal electron states are not treated explicitly but only
implicitly through their density response to an external electrostatic
field, which can be captured in a simple model. Here, we have
explicitly considered the perfect conductor approximation for the
metal response although the method is not strictly limited to this
approximation. In the perfect conductor model the screening length by
the conduction electron is assumed to be zero and the screening charge
resides on a plane. Second, different charge states of the charged
system can be handled directly. Based on the simple perfect conductor
approximation for the metal response, we have implemented this method
in the density functional theory VASP code.

A simple illustration and test of this method is provided by the case
of the Na$^{+}$ ion outside a perfect conductor. The six electrons in
the $p$ semi-core of the ion were included in the calculations.  The
success of our implementation is demonstrated by the excellent
agreement between the calculated Hellman-Feynman force and the
gradient of the interaction energy, even in the region close to the
perfect conductor plane where the electron density of the ion overlaps
with the induced charge density at the perfect conductor
plane. Furthermore, the small polarisability of the ion makes the
calculated interaction energy to be in close agreement to the
analytical result for a point charge outside the perfect conductor
within a supercell of finite size.

Finally, to fulfil our overall aim, we need to include the missing
interactions in our simplified scheme arising from over-lapping
densities and van der Waals interactions between the
ultrathin-insulating film with charged adsorbates and the metal
substrate. We plan to do this by developing a simple parametrised
force field between the metal substrate and the atoms in the adjacent
layer of the film, where the parameters are obtained by fitting the
resulting interactions to DFT calculations of the film (without
adsorbates) over the metal substrate. The presentation of this scheme
and associated results of this model are deferred to a separate
publication.

\ack
This work is supported by the grant (F/00 025/AG) from the Leverhulme
Trust. Mats Persson is grateful for the support from the Swedish
Research Council (VR) and the EU project ARTIST.


\begin{thebibliography}{99}
\bibitem{pauli} The origin of this repulsive interaction is purely
due to Pauli's forces, which arise from the overlapping of the system
and the metal electronic clouds.
\end{thebibliography}
\end{document}